\documentclass[journal=iecred,manuscript=article]{achemso}
\setkeys{acs}{keywords = true}

\usepackage{lineno,hyperref} 

\author{Ravindra S. Ghodake} \affiliation{Academy of Scientific and Innovative
	Research (AcSIR), Ghaziabad 201002 India} \alsoaffiliation{CSIR-National
	Chemical Laboratory, Pune 411008 India}

\author{Ashish V. Orpe} \email{av.orpe.ncl@csir.res.in} \affiliation{Academy of
	Scientific and Innovative Research (AcSIR), Ghaziabad 201002 India}
\alsoaffiliation{CSIR-National Chemical Laboratory, Pune 411008 India}

\title {Surface morphology and shear effect on lubricated granular system}


\begin{document}
	
	\begin{abstract}
		
		Improving the particle surface frictional behavior is of considerable
		importance in several industries so as to manipulate the powder flowability
		and handling. In this work, we have studied the lubrication phenomena in an
		assembly of flowing bulk powder particles in presence of a fine powder
		lubricant. Experiments were conducted in two different shear environments, for
		two different levels of confinement and using three different bulk powder
		particles having distinct surface morphology. The added lubricant particles
		tend to adhere to the surface of bulk particles thereby altering their surface
		morphology. This results in altered inter-particle friction, which typically
		reduces for small addition of lubricant, but increases on its excess addition,
		thereby altering the particle flowability. While this behavior is generally
		understood through previous studies, the current work explores how the effect
		of confinement, shear environment and particle surface characteristics combine
		in a unique manner to provide modified flow behavior for different lubricant
		concentrations. Extended shear duration seems to extend the range of lubricant
		concentrations for improved flowability. On the other hand flow confinement
		seems to limit the extent of lubrication for these range of concentrations.
		Both these opposing effects also exhibit a unique dependence on the particle
		surface morphology and its seeming affinity with the lubricant particle. The
		parameter types and ranges explored comprise most conditions encountered in
		industrial systems, the outcome thereby providing a basis for designing of
		applications handling lubricant-particle mixtures.
		
	\end{abstract}
	
	\section{Introduction} Addition of a small amount of solid, powdered particles
	as a lubricant to an assembly of other particles to alter the inter-particle
	interaction, is routinely practised in pharmaceutical
	applications~\cite{sun11}. One of the objective is to enable smooth ejection of
	tablet from the die, effected by reducing the stickiness between particles as
	well as with the die walls.  Another objective can be to alter the flowability
	of bulk particles by reducing the inter-particle friction or attractive force.
	This effect on flowability has been studied previously to a reasonable extent
	through
	experiments~\cite{navaneethan05,mehrotra2007influence,zhou2010,zhou2010c,zhou2010a,zhou2010b,zhou2011,zhou2011a,mendez12,morin13,sato13,zhou2013,wang17,ghodake2022flow} and to some extent through simulations~\cite{sato13,ketterhagen18} using model, spherical shaped particles while focusing on optimal lubricant concentration vis-a-vis flow characteristics.
	
	For applications involving assembly of bulk particles, quite predominant in
	pharmaceutical processes, the powder lubricant (typically, Magnesium Stearate,
	MgSt) is added in tiny quantities (approximate weight ratio of $O[10^{-2}]$),
	wherein the individual lubricant particles can adhere to the surface of bulk
	particles~\cite{sato13} resulting in altered inter-particle interaction force.
	While this altered force is predominantly due to change in the inter-particle
	cohesion for smaller particles (size around few tens of
	microns)~\cite{zhou2010,zhou2010c,zhou2010a,zhou2010b,zhou2011,zhou2011a,
		zhou2013,wang17}, for large enough particles (size ranging from a few hundred
	microns to millimetres) this altered force is predominantly due to change in
	the inter-particle friction
	~\cite{navaneethan05,morin13,goh2021,ghodake2022flow}.
	
	For smaller sized (few tens of microns) particles, use of mechanical process
	(mechanofusion) has been shown to be an effective method of dry coating
	pharmaceutical powders with MgSt~\cite{zhou2010}. The coating of MgSt reduces
	inter-particle cohesive force resulting in improved powder compressibility and
	flowability, the effect enhancing with increasing host particle
	size~\cite{zhou2010c}. The reduced cohesive force decreases the agglomeration
	strength between particles causing the powders to fluidise efficiently thereby
	improving their aerosolization efficiency required in dry powder
	inhalers\cite{zhou2010a,zhou2010b}. Detailed surface characterisation showed
	that mechanofusion resulted in continuous, uniformly thick coating of MgST on
	the particle surface causing its morphology to change from angular to more
	rounded form~\cite{zhou2011}, the result being improved flowability. The
	effectiveness of coating was also shown to increase with increasing duration of
	mechanofusion process~\cite{zhou2011a}. The increased duration of mixing or
	coating was conjectured to increase lubrication extent from enhanced smearing
	and stretching of MgSt on particle surface~\cite{wang17}. The flowability of
	coated particles was also shown to increase upto an optimal concentration of
	MgSt beyond which it deteriorates~\cite{zhou2013} due to presence of excessive
	lubricant causing disruption of the particle surface.
	
	While the above studies show that predominantly inter-particle cohesive
	interaction gets modified for small particles, in an alternate study it was
	shown that both, frictional as well as cohesive forces can remain dominant for
	small particles~\cite{deng17}. Moreover, they, both, simultaneously decrease
	with increasing surface coverage of bulk particles by lubricant particles
	followed by a slight increase at very large surface coverage. The decrease in
	the interactive force is attributed to systematic change in the contact pattern
	varying from that between bulk particles, through bulk and lubricant particles
	and only between lubricant particles at very large coverage. This variation in
	the type of contacts is also shown to influence the quality and time duration
	of mixing~\cite{deng18}. It is to be noted, however, that this contrasting
	behavior is observed for Silica particles used as a lubricant instead of MgSt,
	which may induce different inter-particle interactions. Similar change in the
	type of particle contacts was also observed in a recent DEM study, but during
	the flow of large (millimetric) sized particles atop an inclined
	chute~\cite{chaudhary26}.
	
	For such large sized particles (few hundreds of microns), maximal increase in
	lubrication, similar to that for fine powders, has been observed previously
	~\cite{morin13}.  The inter-particle friction coefficient measured in terms of
	static angle of repose was shown to decrease to a constant value with increased
	lubricant concentration leading to reduced avalanching time. Similar, improved
	flowability upto a constant value with increased lubricant concentration was
	observed in other studies~\cite{navaneethan05,goh2021}. In our previous
	work~\cite{ghodake2022flow}, it was observed that the friction coefficient
	exhibited non-monotonic dependence (decrease followed by increase) with
	increasing lubricant concentration. The observed bulk behavior essentially
	represents bulk particle surface driven phenomena as observed through
	microscopy imaging~\cite{morin13,sato13,ghodake2022flow}, with negligible
	dependence on bulk particle size~\cite{ghodake2022flow} and the transition from
	a rough, uneven particle surface morphology to a much rounded, smooth surface
	particle due to lubricant coating~\cite{sato13,zhou2011}.  However, the effect
	of the type of bulk particle surface morphology on the flow lubrication is not
	studied in detail, except for few sporadic
	literature~\cite{navaneethan05,zhou2011}, inspite of the direct relevance of
	between the two. Also unknown is the possible effect of the type of flow
	geometry and confinement on the overall lubrication effect. This also entails
	how does the presence of lubricant changes the flow regime from slow, friction
	dominated to fast collision dominated, inertial regime and its sensitivity to
	bulk particle surface morphology and the type of flow environment.
	
	As described in next section, the particles used in this work are millimetric
	in size, essentially larger than those encountered in pharmaceutical powders
	and are flowable and non-sticky (non-cohesive) in nature. Thus, the outcome
	from this work may not be directly applicable to cohesive, non-flowing, micron
	sized powders encountered in practice. However, it will still be of immense
	help to better understand the fundamental quantity, inter-particle friction,
	which governs the flow of granular materials, and whose behavior is of wider
	interest~\cite{midi04}. Typically, altering inter-particle friction either
	requires some kind of chemistry or altering the material, both of which pose
	limitations. Further, DEM simulations allow for studying dry granular systems
	by suitable tuning of inter-particle friction and collisional
	properties~\cite{silbert01,silbert03,orpe19}. But the technique still presents
	limitations in terms of availability of experimental data for a detailed
	comparison in spite of several reported studies~\cite{midi04}. Thus, there is a
	need for the ability to easily modify the inter-particle friction, which seems
	possible by using powdered lubricant. Overall, this should help to advance
	fundamental knowledge about granular materials while adding on to prevalent
	knowledge from existing
	studies~\cite{navaneethan05,mehrotra2007influence,morin13,sato13,wang17,ghodake2022flow,mendez12,ketterhagen18}.
	
	In this work, we try and answer these questions by experimental study of
	lubrication effects on bulk particles of different sizes, types and surface
	morphologies for a given type of lubricant powder.  Further, we seek to
	understand the flow behavior of powder-lubricated materials across two
	different shearing systems: the chute system and the concentric cylinder
	system. These two systems are well studied and researched over the last few
	decades~\cite{midi04,forterre08} so that the flow therein and its dependence on
	various parameters is understood quite well. Each geometry offers a distinct
	environment for controlled shearing of powder materials, providing valuable
	insights into their complex flow dynamics. The systems are so chosen to access
	different types of shear, driving forces and level of confinement. The free
	surface in the inclined chute is unbounded, while the flow in the concentric
	cylinder system remained confined, at least in shear direction. The flow in
	chute is driven by gravity while that in concentric cylinder system is driven
	by external imposed flow (or drag). We expect the type of shear environment to
	have a possible secondary, but distinct effect on the lubrication of bulk
	particles, thereby allowing us to explore the scope of lubrication phenomena
	which will be of relevance in industrial application for estimating the
	lubrication effects.
	
	\section{Materials and Methods} \subsection{Physical properties of powder
		materials}
	
\begin{figure} \centering \includegraphics[width=1.0\textwidth]{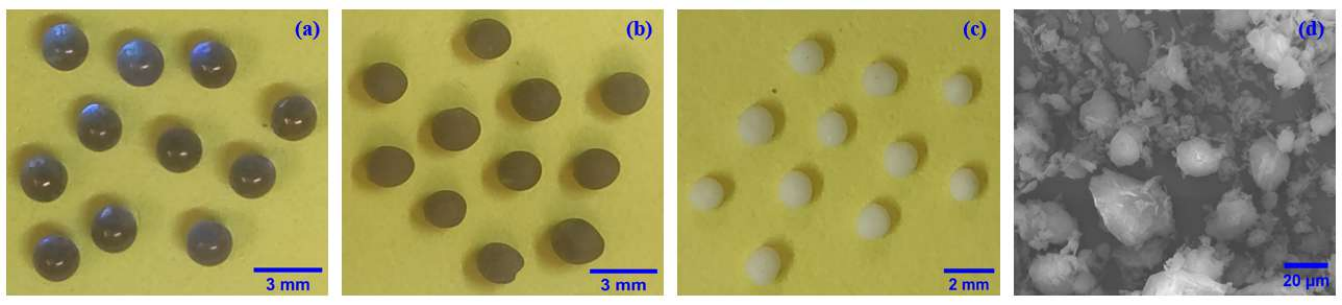}
		\caption{Images of the particles used as bulk and lubricant powder material:
			(a) Glass beads, (b) Mustard seeds, (c) Lactose granules and (d) Magnesium
			Stearate (MgSt). Surface of MgST is imaged using FE-SEM technique, while others
			are imaged using a digital camera.} \label{partimg} \end{figure}

	Three types of nearly spherical particles, shown in fig.~\ref{partimg} and
	detailed in Table~\ref{partchar}, were used as bulk powder materials. Glass
	beads and mustard seeds were sourced commercially while Lactose granule were
	prepared using in-house granulation technique. To prepare the granule, Lactose
	monohydrate and microcrystalline cellulose powder (Loba Chemie, India), were
	first mixed in a 70 : 30 weight ratio. The blend so obtained was mixed with 50
	$wt$\% water to obtain a homogeneous, pasty mass. This paste was extruded
	through 1 mm diameter holes producing cylindrical extrudates with an
	approximate length to diameter (L/D) ratio of 2 : 1. The freshly formed
	extrudates were transferred to a spheronizer comprising a fixed cylindrical
	chamber with a horizontal, grooved plate placed near its bottom which could be
	rotated at a desired speed. The plate was rotated at $1100$ rpm and the grooves
	broke the extrudates into shorter segments.  These segments, through their
	continuous collision and rolling along the chamber side walls, gradually
	rounded into spherical granules.  The granules were removed from the
	spheronizer and dried overnight under ambient conditions before being used in
	lubrication experiments. The final granule size is primarily determined by the
	extruder hole diameter ($1$ mm), the groove design of the spheronizer plate,
	and the rotational speed.  Since the extrusion step fixes the diameter of the
	cylindrical extrude to $1$ mm, the resulting granules exhibit a very narrow
	particle size distribution (PSD), with an average particle diameter of
	approximately $1.0$ mm. It is to be noted that the choice of granulation
	process was incidental as that was available in our group. Though these
	granules can be prepared from any other well known techniques.
	
	Magnesium Stearate (MgSt) powder is used as a lubricant material.  These
	lubricant particles are cohesive, plate-like particles with an equivalent
	sphere volume diameter of $10~\mu$m with $10$\% polydispersity and density of
	$1.0284$ g$/$cm$^{3}$.  These plate-like particles, after mixing with bulk
	particles, tend to spread over latter's surface in the form of a film.  The
	friction between bulk particles is reduced due to contact between these
	lubricant films which have very less friction with respect to each
	other~\cite{miller1985frictional}. However, the reduced frictional property of
	lubricant particles is manifested only on spreading and stretching of
	plate-like particle over another surface, and not for a separate, standalone
	assembly of un-stretched lubricant particles. This behavior is quite well known
	in pharmaceutical powder mixing~\cite{morin13,goh2021,li2014lubricants} wherein
	the lubrication function of MgSt powder is associated with ability of its
	particles to coat host particles.
	
\begin{figure} \centering \includegraphics[width=1.0\textwidth]{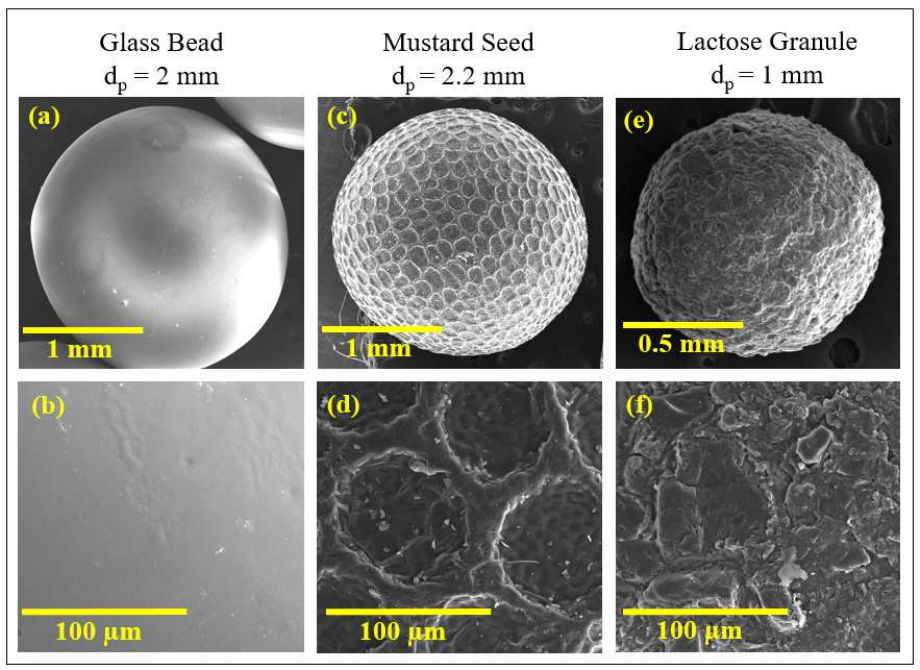}
		\caption{FE-SEM images of bulk powder materials. Top panel represents images at
			particle scale and bottom panel represents magnified view of the particle
			surface. The length scales are provided for ready reference.} \label{fesem1}
	\end{figure}

	The surface morphology of bulk powder particles was imaged using Field Emission
	Scanning Electron Microscopy (FE-SEM). Figure~\ref{fesem1} shows images of the
	surface of each of these particles at two different magnifications. The surface
	of glass bead (see fig.~\ref{fesem1}a and b), seems to be very smooth, with
	minimal asperities unlike the other two, each of which has distinct patterns.
	The surface of mustard seeds (fig.~\ref{fesem1}c and d) have well defined and
	uniformly spread islands which are separated by reasonably wide bands or
	ridges. The images, however, do not reveal if these bands are elevated or
	depressed from the surface of the particle.  The surface of Lactose granules
	(fig.~\ref{fesem1}e and f) shows similarity with that of mustard seeds, though
	the patches are spread non-uniformly, are of different sizes without any
	well-defined bands or ridges in between. Given such uniqueness in the surface
	morphology of each bulk powder particle, the inter-particle contact behavior
	can be expected to exhibit different characteristics.
	
	\begin{table} \caption{Physical characteristics of powdered materials}
		\label{partchar} \begin{tabular}{lcc} \hline Particle Name  & Particle diameter
			($d_p$)mm & Particle density ($\rho_p$) g/cc \\ \hline Glass bead & $2 \pm
			0.15$ & $2.5$   \\ Mustard seed & $2.2 \pm 0.15$ & $0.78$ \\ Lactose granule &
			$1.0 \pm 0.1$ & $1.8$  \\ \hline \end{tabular} \end{table}
	
	\subsection{Experimental System} \label{sec:expsys}
	
\begin{figure} \centering \includegraphics[width=1.0\textwidth]{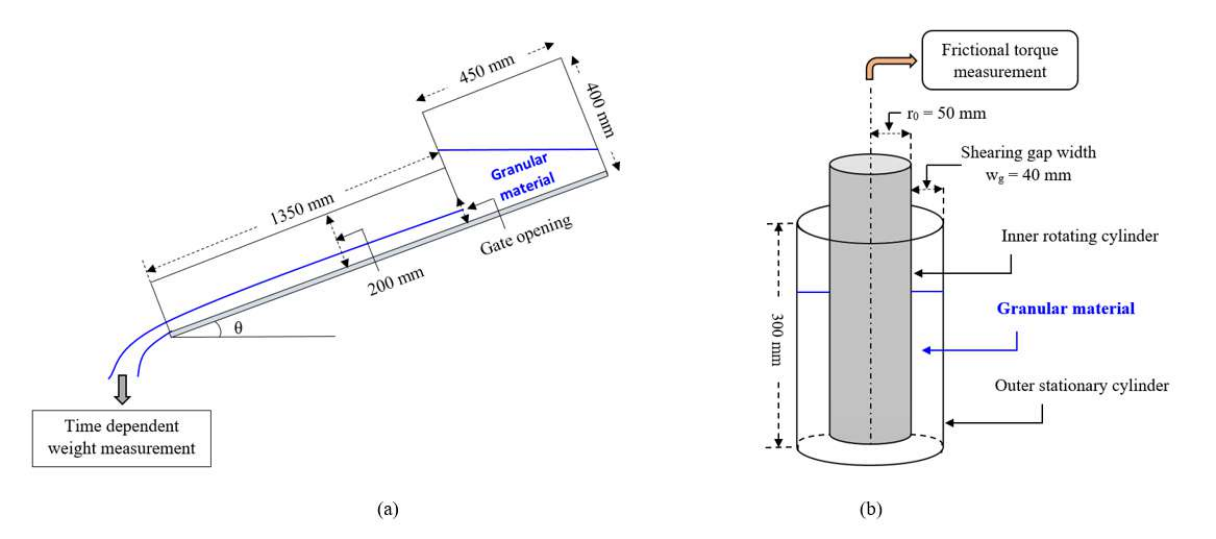}
		\caption{Schematic of different experimental systems with relevant dimensions
			and flow directions. (a) Side view of chute system and (b) Perspective view of
			concentric cylinder system.} \label{expsys} \end{figure}

	Experiments are carried out in two flow geometries: the inclined chute system
	and concentric cylinder assembly.  The chute system comprises two sections as
	shown in fig.~\ref{expsys}a, chute inclined at an angle ($\theta$) and a
	hopper. The hopper is $450$ mm in length, $400$ mm in height and $50$ mm in
	width and is located at the top end of the system and functions as a reservoir
	for the powder material. The remaining parts of system, of length $1350$ mm,
	width $50$ mm and height $200$ mm, serves as a chute. The powder material flows
	on to chute surface from the reservoir through a gap of prescribed height (h)
	which is regulated using a manually operated gate. The system side walls are
	constructed out of stainless steel (SS$316$), except for a transparent acrylic
	window of length $600$ mm and height $200$ mm, located downstream ($600$ mm
	from chute end) for imaging the flow. The base of the chute (of width $50$ mm)
	is made from glass and is roughened by gluing glass beads of desired diameter
	($d_p$) on it. The material flowing out of the chute near the lower end is
	collected onto the weighing balance to record time dependent flow rate.
	
	For each type of powder material, the chute was maintained at a specified angle
	with the horizontal known as inclination angle ($\theta$). Generally, a
	progressive increase in the chute inclination angles leads to faster flows and
	consequently reduced time duration of steady state flow. As described later on,
	the higher duration of steady flow rate is desirable for acquiring reasonable
	number of flow images to get better statistics for time averaged flow
	velocities. On the other hand, decreasing the angles led to slower flows with
	increasing probability of flow intermittency. Considering the significant
	differences in size and type of powder material used in the work, the chute
	inclination angle was finalised while striking a balance between flow duration
	and flow smoothness. However, achieving this balance was not necessarily
	achieved at the same inclination angle for each type of particle size and type.
	The optimisation process involved finding the inclination angles that provided
	a reasonable compromise between the duration of steady flow needed for data
	collection and the smoothness of the flow, considering the specific
	characteristics of each type of powder material.
	
	The schematic of the concentric cylinder system employed in this work is shown
	in fig.~\ref{expsys}b. The system consists of an inner solid cylinder of radius
	$50$ mm and height $300$ mm located within the outer hollow cylinder (closed at
	bottom and open at top) of radius of $90$ mm and also of height $300$ mm, both
	constructed from stainless steel. The annular gap between the two cylinders,
	measuring $40$ mm, is filled with powder material up to a height of $180$ mm.
	The material within the annular gap is sheared by rotating the inner cylinder
	at a desired angular speed and in a desired direction while the external
	cylinder is held stationary. The torque exerted on the sheared powder material
	is measured using a spring-detector system connected to the inner rotating
	cylinder. To prevent material slippage, the surface of the rotating cylinder
	and the inner surface of the outer stationary cylinder were roughened by
	adhering a single layer of glass beads of prescribed size. The vertical gap
	near the bottom between the two cylinders is less than 0.5 mm, small enough for
	bulk particles to get trapped and cause breakage or jamming leading to
	incorrect torque measurements. While uncoated lubricant particles may
	accumulate in this gap, their smaller size ($1$\% of gap size) and smaller
	amount (see weight ratio mentioned in next paragraph) is unlikely to create any
	operational issues.

	\subsection{Experimental methodology} \label{sec:method}
	
	Experiments were conducted to study the flow of a mixture of bulk powder
	material and MgSt lubricant in the two systems mentioned earlier.  For every
	experiment, a pre-defined, but very small quantity of MgSt powder was mixed
	with each of the three bulk powder materials. The mixing was carried out in an
	open plastic tank (diameter $185$ mm and height $100$ mm) fitted with an
	overhead stirrer, for a duration of $5$ minutes. The rotational speed of the
	stirrer was set at a speed of $30$ revolutions per minute (rpm) throughout. The
	mixing procedure, including mixing time and stirrer speed, was kept constant
	across all experiments. The weight ratio of MgSt to powder material was varied
	between O ($10^-6$) and O ($10^-2$) across all experiments. The maximum value
	used is comparable to that used previously for lubrication of aerosol powders
	~\cite{zhou2010a,zhou2010,zhou2010b,zhou2010c,zhou2011,zhou2011a,zhou2013} in
	the size range of $1-100$ $\mu m$. Further, it also corresponds to a volume
	ratio of O ($10^{-2}$) and a number ratio of O ($10^{4}$). For a given weight
	ratio, the lubricant concentration was defined $c_l$ = $M_l/A_p$, where $M_l$
	is the total mass of lubricant, $A_p$ = n $\pi$ $d_{p}^2$ is the total surface
	area of particles, $d_p$ is particle diameter and n is the number of particles.
	The value of n is determined as ($6/\pi$ $d_{p}^3$) ($M_p$/$\rho_p$), where
	$M_p$ is total mass of particles, $\rho_p$ is the particle density and
	$M_p$/$\rho_p$ is the total particle volume.
	
	In the chute flow, the inclination angle was fixed at $24$ degrees for glass
	beads, $26$ degrees for mustard seeds and $32$ degrees for lactose granules.
	The chute (glass) base was roughened with $1$ mm glass beads for flow of
	lactose granules, while $2$ mm glass beads were used as surface roughness
	during flow of glass beads and mustard seeds. For each experiment, the hopper
	was filled up to $30$\% to $60$\% of its total volume with mixture of lubricant
	and one type of primary powder material while maintaining the hopper gate
	closed. The volume filled in the hopper was dependent on the size and type of
	powder material used, but was adequate to ensure steady flow for a reasonable
	duration in the experiment. The gate was opened manually to allow flow of
	lubricant-powder mixture down the chute surface. The gate opening was
	maintained at $30$ mm throughout to allow free flow for all materials. The
	material exiting the chute was collected directly on a weighing scale to obtain
	its time dependent behavior. A small time duration window (4 s to 20 s)
	exhibited linear variation of mass collected with time which provided steady
	state mass flow rate ($m_{c}$). The required time duration window was dependent
	on the material used and the chute inclination angle employed. Due to the
	varying flowability characteristics of the different powder materials, the
	values of chute inclination angles, base roughness were set differently to
	achieve steady and uniform flow rate in each case.
	
	In concentric cylinder assembly, the powder-lubricant mixture was introduced
	into the gap between two cylinders up to a height of 180 mm. The material was
	sheared by rotating the inner cylinder at a constant angular speed of 5
	revolutions per minute (rpm) for a duration of 60 minutes while keeping the
	outer cylinder stationary.  This corresponds to apparent shear rate of $7.9 $
	s$^{-1}$ employed in the shear gap of 40 mm (assuming linear velocity profile)
	with inner rotating cylinder of radius 50 mm.  Over the entire duration of 60
	minutes, the measured torque value oscillates about the mean with no apparent
	and perceivable dependence on shearing time (not shown). This behavior is
	observed for all three particles and for all lubricant concentrations studied.
	The fluctuations about the mean possibly represent the non-uniformity of the
	lubricant coating over the particles which, may alter with time. Perhaps, the
	lubricant is spreading over particles with shearing time, but it is not exactly
	similar for all particles due to possible non-uniform coating created
	initially. The reported values of torque represent the average over the entire
	60 minute duration. The fill height, rotational speed of the inner cylinder and
	duration of material shearing was maintained constant for all the materials
	studied to ensure uniform conditions throughout the experimental trials. During
	the shearing process, torque readings were recorded at intervals of 60 seconds.
	The average torque for a given material and lubricant concentration was
	obtained as the mean of the near-constant torque values over the entire
	duration of the experiment.
	
	Apart from the bulk measurements (flow rates and torque values) mentioned
	above, flow visualisation studies were conducted to understand the microscopic
	details of the flowing material. However, this could be carried out only for
	transparent and reflective glass beads. Identification of positions of other
	particles was difficult due to their opacity and lack of reflection which gets
	accentuated with lubricant coating.
	
\begin{figure} \centering \includegraphics[width=1.0\textwidth]{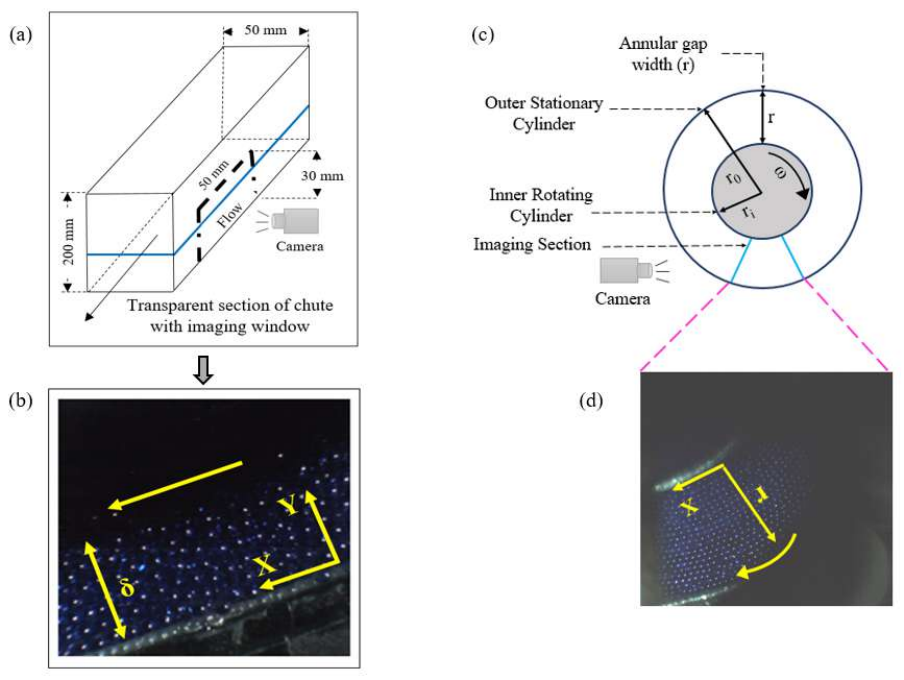}
		\caption{Schematic representation of the imaging methodology implemented in
			experiments. (a) Transparent section of the inclined chute system with the
			imaging window (b) Sample image of flowing material on the inclined plane at
			the side wall (c) Top view of the concentric cylinder system (d) Image of the
			flowing material at the top, free surface of the rotating material. All the
			images were captured using a high speed camera (see text for details).}
		\label{expmeth} \end{figure}

	The steady flow of glass beads near the side walls was recorded using the IDT
	Y4 high speed camera. The camera was positioned orthogonal to the chute side
	wall and images were recorded in the observation window of length $30d_p$ and
	height $20d_p$ (see fig.~\ref{expmeth}a). This observation window was located
	at a distance $300$ mm away from the chute exit. The observation window was
	positioned downstream of the chute entrance to ensure fully developed powder
	flow and sufficiently away from the chute end to avoid exit effects. The camera
	was operated at a frame rate of $1500$ frames per second, resulting in the
	acquisition of approximately $6000$ images with resolution of $1024 \times
	1024$ pixels. A typical image of flowing glass bead is shown in
	fig.~\ref{expmeth}b. The particle centers in each image were obtained using IDL
	(Interactive Data Language) centroid algorithm available as open source
	software~\cite{idl}. The instantaneous particle velocities in $x-y$ plane were
	measured from the particle positions in successive images and the time delay
	between successive positions. Here, ``$x$'' represents flow direction and
	``$y$'' represents direction transverse to flow.
	
	The entire flowing layer was divided into horizontal bins of thickness ($w =
	d_p$) and length ($l = 20 d_p$), oriented parallel to the chute surface. The
	flow direction velocity component ($v_{xc}$) and transverse direction velocity
	component ($v_{yc}$) for each bin was obtained as an average over all particle
	identified within the bin and across all images. This resulted in profiles of
	x-direction and y-direction velocities with distance $y$ from the chute
	surface. Typically, the number of centroids (i.e. particles) detected per bin
	per image was about $20$ close to the base and decreased rapidly close to the
	free surface. The free surface location was, then, defined as the bin within
	which not less than $5$ particles were detected per image. The distance of the
	free surface from the chute base was defined as flowing layer thickness
	($\delta$). Given that the measurements were carried out at the wall, we have
	calculated the particle area fraction ($\phi_{ac}$) in the flowing layer using
	the projected area (circle) of glass bead. The particle area fraction
	($\phi_{ac}$) in the flowing layer is obtained as $n \times A_{p} / A_{box}$,
	where $A_p = \pi$ $d_p^2/4$ is the area of glass bead, $n$ is the average
	number of glass beads (or centroids) counted per image and $A_{box} = \delta
	\times l$ is the area of flowing layer of length $l$ and thickness $\delta$. 
	Within the visualisation zone itself, images were taken at three different
	locations in x-direction, about $20d_p$ apart from each other. The velocity
	profiles for these three locations were nearly the same suggesting of a fully
	developed flow. Experiments were performed for $8$ to $10$ different lubricant
	concentrations.
	
	The flow visualisation measurements in concentric cylinder system comprised
	imaging of the flow at the free surface to obtain velocity profiles (see
	fig.~\ref{expmeth}c).  The illuminated flow region was reflected from a
	high-quality and $50$ mm $\times$ $50$ mm sized mirror placed at an angle to
	the free surface and recorded by a high-speed camera (IDT Y4) positioned in
	line with the reflected rays from the mirror. This arrangement, shown in
	fig.~\ref{expmeth}d was used to visualize the flow since direct imaging from
	above and normal to flow was not possible due to specific mechanical design of
	the system.  The camera is operated at a frame rate of $120$ frames per second,
	capturing approximately $6000$ images over the course of the experiment, each
	with a resolution of $1024 \times 1024$ pixels. The images were recorded over
	an observation window measuring of size $20 d_p \times 20 d_p$.  They were
	analysed to obtain the velocity profiles in the flow direction.  The
	instantaneous particle velocities in the $x-r$ plane were obtained from the
	particle positions in successive images and the time delay between these
	positions. The entire flowing layer was divided into horizontal bins of
	thickness ($w = 1 d_p$) and length ($l = 5 d_p$), oriented parallel to the wall
	of cylinder. The velocity in flow direction, for each bin was computed as the
	average of all particles identified within the bin across all images, resulting
	in profiles of x-direction velocities ($v_{xcc}$).  The area fraction
	($\phi_{acc}$) was calculated using $A_{box} = 5 d_{p} \times 20 d_{p}$ with
	remaining methodology the same as used in chute system.  All the experiments
	were carried out at a constant temperature of $23$ degC. After every
	experiment, the surfaces of the geometries were thoroughly cleaned to
	completely remove the lubricant particles that may have accumulated from the
	previous experimental run.
	
	\section{Result and Discussion}
	
	We, first, discuss the observed change in the surface morphology of the bulk
	powder particles, due to addition of lubricant particles, using visualisation
	technique followed by quantifying the resulting outcome in the form of a
	measurable quantity. The change in the flow behavior due to change in the
	surface morphology of the bulk powder particles is discussed in further two
	sections for different flow geometries and particle types.
	
	\subsection{Effective friction behavior}
	
\begin{figure} \centering \includegraphics[width=1.0\textwidth]{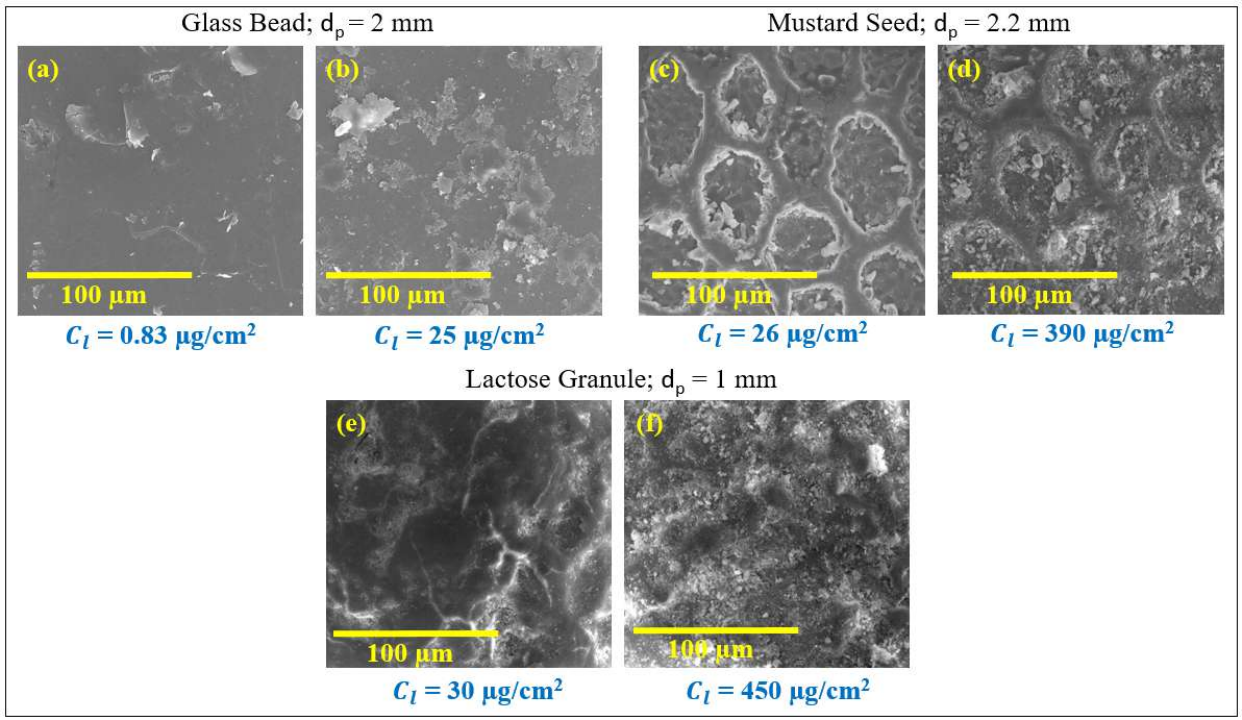}
		\caption{FE-SEM images of each type of bulk powder particle mixed with
			lubricant having two different concentrations ($c_l$).  The scales are provided
			for ready reference.} \label{fesem2} \end{figure}

	The MgSt powder is slightly cohesive in nature and quickly adheres to the
	surface of bulk powder particles during mixing. The mixing process may also
	cause shearing of the lubricant particles thereby stretching and spreading them
	over the bulk particle
	surface~\cite{sato13,morin13,ghodake2022flow,zhou2011,wang17}. The surface of
	the particles, after mixing with lubricant powder, was visualised using FE-SEM
	technique. Figure~\ref{fesem2} shows magnified images of each of the bulk
	powder particles mixed with lubricant powder at two different concentrations.
	
	For smooth surfaced glass beads, the added lubricant accumulates on the surface
	resulting in few number of sparse patches at lower concentrations (see
	fig.~\ref{fesem2}a), which increases at higher lubricant concentrations
	(fig.~\ref{fesem2}b). Given the lack of any obvious asperities on glass surface
	(see fig.~\ref{fesem1}a and b), the only role for lubricant in this case is
	accumulation on the surface. However, the thickness of these patches remains
	uncontrollable and unmeasurable. The presence of these patches during the
	contacts between bulk particles is expected to reduce their friction given the
	near frictionless nature of MgSt particles. In case of mustard seeds, the
	lubricant seems to accumulate at the periphery of the individual islands for
	lower lubricant concentration, while it accumulates on the island itself at
	higher concentrations (see fig.~\ref{fesem2}c and d). The surface roughness of
	mustard seeds, arising out of presence of islands and adjoining bands, does not
	seem to be smoothened out due to added lubricant. The inherent frictional
	behavior due to this roughness is, then expected to be retained but may be
	offset to some extent due to lubricant patches on the surface of island which
	will be present at inter-particle contacts. The surface of lactose granules, on
	addition of a small amount of lubricant, appears smoother (see
	fig.~\ref{fesem2}e) compared to the jagged morphology in the absence of
	lubricant (fig.~\ref{fesem1}f). This is expected to reduce the inter-particle
	friction behavior.  Similar smoothening and eventual rounding of the Lactose
	particles on addition of MgSt lubricant has been noticed
	previously~\cite{morin13,zhou2011,wang17}. While the surface morphology of
	Lactose granules looks similar to that of mustard seeds, the degree of
	smoothening is different suggesting different affinity of lubricant particles
	with the type of bulk particle surface. Again, at higher concentrations, the
	surface seems to become rougher due to coating of more lubricant particles (see
	fig.~\ref{fesem2}f).
	
	Figures \ref{fesem1} and \ref{fesem2} indicate that the presence of lubricant
	particles alters the surface morphology of the bulk particles. The extent of
	coating does not seem to be uniform for given particle surface, thereby
	requiring quantification of lubrication effect using an average quantity as
	detailed below rather than particles surface characterisation. Moreover, the
	coating is also not uniform across different types of particle surfaces
	suggesting varying affinity with respect to a type of particle surface.
	
	This altered morphology will affect the inter-particle frictional behavior as
	mentioned above, though quantifying this at an individual particle level is
	quite difficult using any of the existing techniques given the large
	(millimetric) size of bulk particles. Moreover, measuring at the individual
	particle level is not advisable given the expected non-uniform distribution
	across all the particles during mixing, thereby, requiring a very large sample
	for any meaningful average. The estimate of this change in the frictional
	behavior, then needs to be quantified using a bulk quantity, say static angle
	of repose, as was done in earlier work~\cite{ghodake2022flow}. The tangent of
	this angle ($\theta_{r}$) with respect to the horizontal can be interpreted as
	the effective friction coefficient ($\mu_{eff}$ ) and has been used in previous
	studies for particle sizes ranging from a few tens of
	microns~\cite{zhou2013,wang17}, few hundred microns~\cite{morin13} or few
	millimeters\cite{pou99}. It is to be noted that this measurement represents a
	bulk property and can also be dependent on various characteristics like
	particle shape, roughness, surface energy, electrostatic effect amongst and
	reflects average value over several particles. However, all these
	characteristics are expected to remain the same for a given particle type. 
	Further, this effective friction is different than the macroscopic friction
	encountered during the flow of material as discussed
	previously\cite{jop06},\cite{forterre08}, which is dependent on inertial state
	and underlying collision dynamics. Decoupling this macroscopic friction from
	effective friction is not possible, since the inertial state directly results
	from collision dynamics, which itself is dependent on lubricant concentration
	as discussed later.
	
\begin{figure} \centering \includegraphics[width=0.8\textwidth]{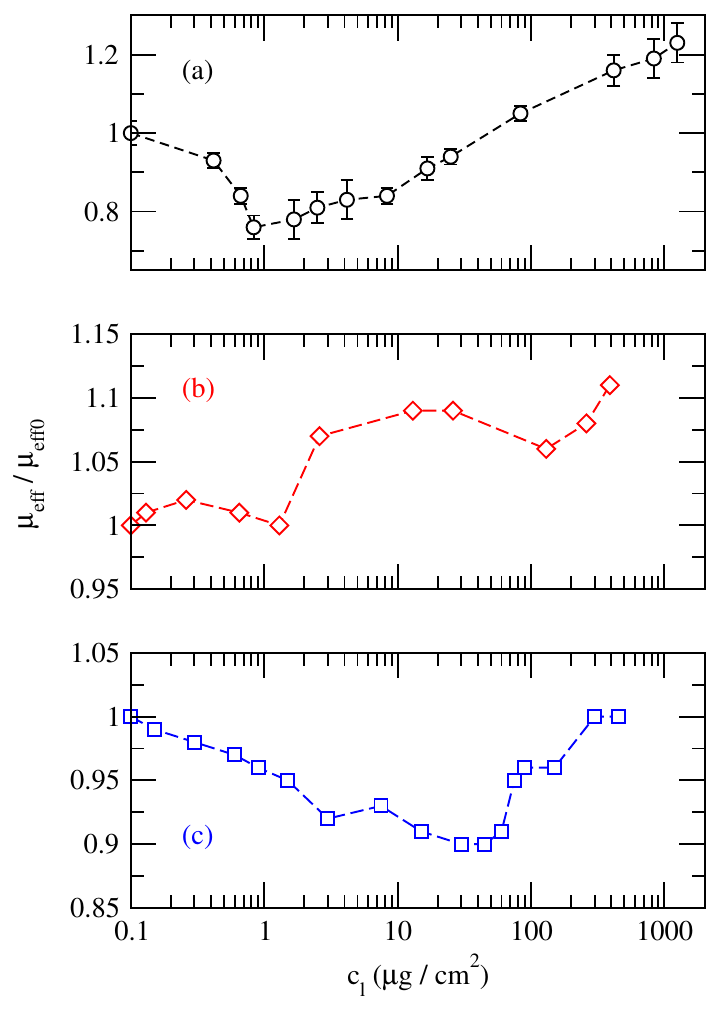}
		\caption{Variation of normalised effective friction coefficient ($\mu_{eff}$ /
			$\mu_{eff0}$) with lubricant concentration ($c_{l}$) for (a) Glass beads (b)
			Mustard seeds (c) Lactose granules.} \label{friction} \end{figure}

	For measuring the static angle of repose, the bulk particles, with or without
	mixed lubricant, were slowly poured in a rectangular cell of height $21$ cm,
	length $48$ cm, and width $3.5$ cm to form a heap.  The length and height of
	the system is sufficiently large to avoid end effects. The entire system is
	constructed from transparent acrylic to facilitate imaging of the heap formed
	by the poured material. The static heap, after the pouring was complete, was
	imaged with a digital camera placed laterally and orthogonal to the side wall
	of the rectangular cell, and the image was analysed to determine the
	inclination of the free surface relative to the horizontal ($\theta_r$). A
	central, nearly flat, free surface region of the heap, approximately $20$ cm in
	length, was selected for analysis. The variation of normalised effective
	friction ($\mu_{eff}$ / $\mu_{eff0}$) with lubricant concentration ($c_l$) for
	all the three types of particles is presented in fig.~\ref{friction}.  Here,
	$\mu_{eff0}$ corresponds to the value of friction in the absence of lubricant
	($c_l$ = 0.0 $\mu$g$/$cm$^2$), representing the base case scenario, which
	cannot be shown on a logarithmic scale. The values of $\mu_{eff0}$ are 0.42,
	0.47 and 0.52, respectively, for glass beads, mustard seeds and lactose
	granules. The smallest value of $c_{l}$ represents the minimum possible content
	of lubricant that can be measured accurately and added to the bulk powder
	material. We discuss some of the important features from fig.~\ref{friction}
	next.
	
	For high enough lubricant content ($c_{l}$) and for all the particles studied,
	the value of normalised effective friction coefficient
	($\mu_{eff}$/$\mu_{eff0}$) is greater than $``1"$.  This suggests that the
	frictional behavior at these values of $c_{l}$ actually worsens than the base
	case, thereby putting an upper limit on the addition of lubricant to the bulk
	powder particles. For concentration range below this upper threshold, all the
	particles except mustard seeds exhibit a minimum in the value of effective
	friction coefficient at intermediate lubricant content. The minimum represents
	maximum flowability possible relative to the base case due to addition of
	lubricant powder. The values of $c_{l}$ corresponding to maximum as well as
	minimum flowability is different for different bulk powder particles. This
	suggests a strong dependence on the type of surface morphology and the affinity
	of lubricant with the bulk particle surface. This may result in coating pattern
	which is different for different particle surfaces, even for the same value of
	$c_{l}$, thereby influencing specific behavior.  At the same time, it is
	possible that the differences may also arise due to unique physical properties
	of individual material, its bulk density, contact mechanics and packing
	structure, all of which may affect the friction behavior. However, it is not
	possible to quantify these effects by any physical measurements.  In essence,
	the normalised lubricant concentration ($c_{l}$) used in subsequent discussion
	may not be the most appropriate measure that can be used for comparing the
	behavior of particles with different surface properties, since unique
	properties of a specific surface plays vital role. It is, however, being used
	to represent average quantity of lubricant powder available for coating a unit
	surface area.
	
	The decrease in the value of normalised effective friction coefficient for a
	specific material can be attributed to the presence of lubricant particles on
	the bulk particle surface which interact during inter-particle contact. For
	large enough particles studied over here, the decrease is attributed to reduce
	friction during contact due to presence of near frictionless lubricant
	particles~\cite{morin13,ghodake2022flow}. For very fine powders, the presence
	of same lubricant particles is shown to reduce inter-particle cohesive
	attraction resulting in reduced angle of repose or effective
	friction~\cite{zhou2013,wang17}. Clearly, progressive increase in lubricant
	concentration causes increasing lubricant presence resulting in progressive
	reduction in effective friction.
	
	For glass beads, the decrease may be attributed to mere presence of lubricant
	particle on surface. For Lactose granules and mustard seeds, an additional
	effect may be due to smoothening of their irregular surface, thereby allowing
	better relative motion between particles, though this effect could not be
	verified through independent measurements. However, such smoothening effect is
	not unusual and has been observed previously, wherein the presence of lubricant
	resulted in better rounding of the angular shaped Lactose
	particles~\cite{zhou2011} or Lactose granules~\cite{morin13}. The contribution
	of both, frictionless lubricant particles and surface smoothening, perhaps,
	provides improved reduction in effective friction and over a wider range of
	lubricant concentration. For Lactose granules, the reduction occurs over a very
	wide lubricant concentration range, $0.1 \mu$g$/$cm$^{2} - 50 \mu$g$/$cm$^{2}$
	(see fig. 6c), which is in the similar range to that reported previously ($1.96
	\mu$g$/$cm$^{2} - 29.4 \mu$g$/$cm$^{2}$) for lubrication of fine Lactose
	powders for aerosol
	applications~\cite{zhou2010a,zhou2010,zhou2010b,zhou2010c,zhou2011,zhou2011a,zhou2013}. The size of the powders in those studies, however, ranged between $1 - 100 \mu$m compared to $1$ mm sized Lactose particles used over here.
	
	Surprisingly, the friction for mustard particles does not decrease for any
	lubricant content, the reasons for which are unclear. While one may attribute
	this lack of friction reduction to the incompatibility of surface energy
	between lubricant and bulk surface, we believe it has to do with improper
	coating of lubricant arising from small duration of mixing and flow since the
	same mustard particles exhibits better flowability in concentric cylinder
	system (shown later) for the same range of lubricant content explored here.
	
	Apart from mustard seeds, for other two particles, there is an upper (optimal)
	limit of lubricant content above which the friction coefficient starts to
	increase (or flowability worsens). Similar, though not exactly same, optimality
	for lubricant content has also been reported previously for MgST coated fine
	powders~\cite{zhou2013}, Silica coated fine powders~\cite{dave17,dave23}, DEM
	simulations~\cite{deng17} and MgSt coated glass bead flow in chute
	system~\cite{ghodake2022flow}. In the current work, on substantial addition of
	lubricant the friction coefficient actually increases than that observed for
	the base case. We think that this increase in friction behavior may be
	attributed to excess accumulation of MgSt particles on the bulk surface
	resulting in an irregular surface. While this cannot be confirmed by images in
	fig.~\ref{fesem2}, similar roughening of particle surface due to excess
	lubricant has been reported previously for lactose particles mixed with calcium
	stearate as lubricant~\cite{morin13}, resulting in reduced flowability. At the
	same time, this behavior is somewhat different than other studies wherein the
	friction coefficient remained constant near the minimum~\cite{zhou2010,morin13}
	or the flowability did not exhibit further
	improvement~\cite{navaneethan05,goh2021} for any further increase in the
	lubricant content. We attribute these differences to the different types of
	particles used in both studies, which can result in different contact mechanics
	and packing fraction resulting in different frictional dependence on lubricant
	content. In the current study, the minimum in the friction coefficient occurs
	at different $c_{l}$ for particles with different morphologies. In case of
	mustard particles, the only effect of lubricant is to worsen the frictional
	properties at all concentrations.

	\subsection{Lubricant effect in different flow systems}
	\label{sec:lubr-effect-geom}
	
	The influence of the type of flow on lubrication behavior is discussed in this
	section. The flow of $2$ mm glass beads is analysed in two different systems to
	primarily ascertain the effect of the type of shear and geometry. The same
	analysis is, then, extended in the next section to ascertain the effect of the
	surface morphology of bulk particles with respect to specific flow geometry.
	
\begin{figure} \centering \includegraphics[width=0.8\textwidth]{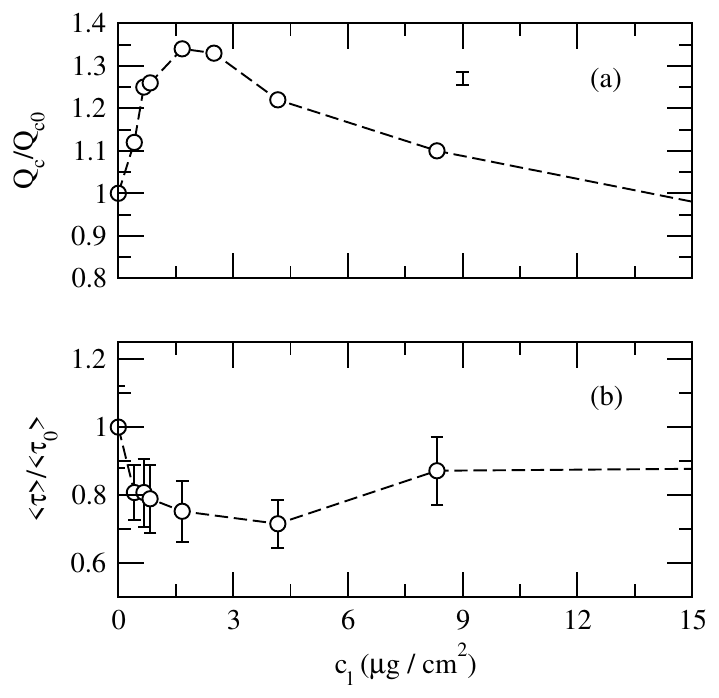}
		\caption{Variation of (a) normalised flow rate ($Q_c$/$Q_{c0}$) in chute system
			and (b) normalised torque ($\tau$/$\tau_0$) in concentric cylinder assembly,
			with lubricant concentration ($c_{l}$) for $2$ mm glass beads. Single error bar
			in (a) represents the maximum deviation for the flow rate data.}
		\label{bulkflowsys} \end{figure}

	Figure~\ref{bulkflowsys} represents the flow behavior of glass beads in both,
	the inclined chute and concentric cylinder systems for varying lubricant
	concentrations. The behavior is shown in terms of variation of normalised
	volumetric flow rate in chute system ($Q_c$/$Q_{c0}$) and normalised torque
	($\tau$/$\tau_0$) in concentric cylinder system with lubricant concentration
	($c_{l}$). Here $Q_{c0}$ and $\tau_{0}$ are the respective values in the
	absence of lubricant, referred as the base case. A non-monotonic dependence is
	seen with respect to lubricant content in both the systems. The increase in
	flow rate values (fig.~\ref{bulkflowsys}a) to a maximum and decrease in the
	torque values (fig.~\ref{bulkflowsys}b) to a minimum for an intermediate value
	of $c_{l}$ corresponds to improved flowability on account of reduced friction
	(see fig.~\ref{friction}a). The occurrence of minimum or maximum suggests
	largest possible improvement in the flowability.  A substantial reduction in
	the flow rates and increased torques values at higher lubricant concentration
	represents reduced flowability due to increased friction, which itself exhibits
	a non-monotonic dependence on $c_{l}$ (see fig.~\ref{friction}a).
	Interestingly, the relative change with respect to the base case is different
	in different flow systems.  Further, the rate of increase (or decrease) from
	base case to maximum (or minimum) is also widely different for different flow
	system. This suggests an additional, independent effect of the flow system
	superimposing on the lubricant effect. We, next, delve into each system
	separately.
	
\begin{figure} \centering \includegraphics[width=1.0\textwidth]{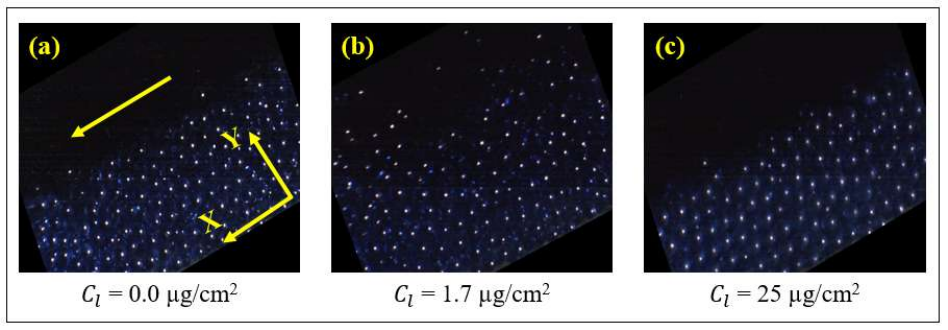}
		\caption{Images of the flowing layer of $2$ mm size glass beads on a chute
			system for three different lubricant concentrations. Chute was maintained at
			angle of $24$ degrees to the horizontal.} \label{chuteimages} \end{figure}

	The flow rate on an inclined plane (or chute) for a fixed particle type is
	typically governed by the inclination angle and inlet gate
	opening~\cite{midi04}, which are held constant in this study. However, instead
	of a single steady flow rate, a range of flow rates are encountered with
	varying $c_{l}$.  The qualitative nature of the flow captured using a digital
	camera is shown in fig.~\ref{chuteimages} during the flow of $2$ mm size glass
	beads at three different lubricant concentrations. All the images were captured
	near the side wall. Fig.~\ref{chuteimages}a, b and c, respectively, depict the
	flow of particles in absence of lubricant, at intermediate lubricant content
	representing maximum flow rate and for a very high lubricant content. The flow
	in absence of lubricant is reasonably packed (fig.~\ref{chuteimages}a), that at
	intermediate lubricant concentration appears expanded (fig.~\ref{chuteimages}b)
	while a densely packed state is observed for high lubricant concentration
	(fig.~\ref{chuteimages}c).
	
	\begin{figure} \centering \includegraphics[width=0.8\textwidth]{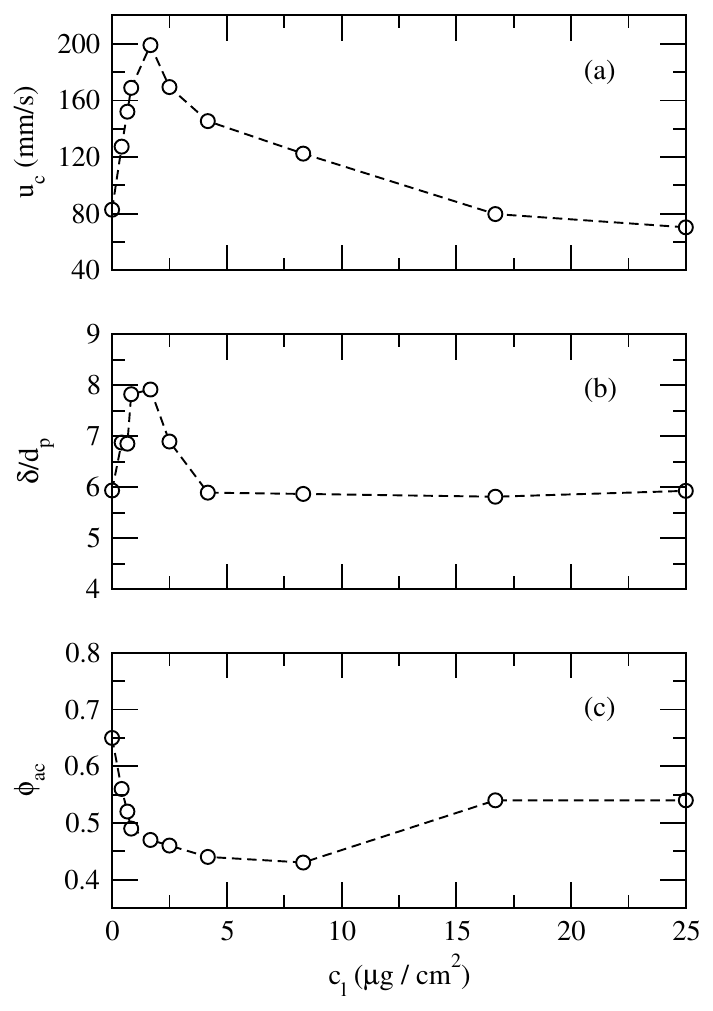}
		\caption{Variation of (a) depth average velocity ($u_c$) of particles in the
			flowing layer over an inclined chute system (b) flowing layer thickness
			($\delta$/$d_p$) and (c) area fraction ($\phi_{ac}$) measured near the walls
			with lubricant concentration ($c_l$) for $2$ mm glass beads. Chute was
			maintained at an angle of $24$ degrees to the horizontal. Error bars are
			smaller than the symbol size and hence not shown explicitly.}
		\label{chuteflowdetails} \end{figure}
	
	We now use the data on mean velocities, layer thicknesses and packing fractions
	to ascertain the predominant factor influencing non-monotonic flow behavior.
	The volumetric flow rate can be expressed as $Q_{c} = u_c$ $\delta$ $\phi_{vc}
	W_{c}$, where $u_c$ = $<v_{xc}>$ is the mean velocity in the flowing layer,
	$W_{c}$ is the width of the flowing layer, $\delta$ is the thickness of the
	flowing layer, and $\phi_{vc}$ is the mean particle volume fraction in the
	flowing layer. Here, $<>$ represents the average over the entire flowing layer
	thickness.  Since $W_{c}$ is kept constant, variations in the flow rate will
	depend on changes in the other three variables as the lubricant concentration
	varies. Further, as all measurements were conducted near the side walls, we
	report the variation in particle area fraction ($\phi_{ac}$) instead of volume
	fraction ($\phi_{vc}$).
	
	The variation of the mean velocity ($u_c$) with lubricant concentration is
	illustrated in fig.~\ref{chuteflowdetails}a.  The mean velocity variation
	nearly mimics the non-monotonic trend for the volumetric flow rate shown in
	fig.~\ref{bulkflowsys}a. The decrease in inter-particle friction coefficient
	due to increased lubricant content leads to a faster flow direction velocity.
	The velocity magnitude increases rapidly for low enough lubricant content. This
	results in enhanced collisions accompanied by lesser dissipation of particle
	kinetic energy causing expansion of the flowing layer. The layer thickness
	($\delta$), thus, shows a rapid increase for lower values of $c_{l}$ (see
	fig.~\ref{chuteflowdetails}b). The same number of particles now occupy a larger
	space resulting in fast reduction in the area fraction values (see
	fig.~\ref{chuteflowdetails}c). The combined effect of these variations is the
	increase in the flow rate for small enough lubricant contents. Exact reverse
	behavior occurs at large enough lubricant contents. Here, additional coating
	(see fig.~\ref{fesem2}b) seems to increase the dissipation during particle
	collisions, thereby slowing down the flow while nullifying any advantage due to
	reduced friction. It is to be noted that the lubricant particles are also quite
	soft (malleable) in nature. The flow velocity magnitude, thus, shows a slow
	decrease with increase lubricant content (see fig.~\ref{chuteflowdetails}a).
	The dissipation leads to flowing layer contraction causing a slow decrease in
	flowing layer thickness (fig.~\ref{chuteflowdetails}b) and eventually increased
	area fraction (fig.~\ref{chuteflowdetails}c) due to same number of particles
	packed in lesser area. The combined effects is, then, to reduce the magnitude
	of flow rate.  The overall behaviour of layer thickness and area fractions are
	quantitative representations of what is observed for flow behavior shown in
	fig.~\ref{chuteimages}.
	
	The non-monotonic behavior of flow rate in chute flow can be summarised as,
	reduced inter-particle friction for intermediate concentrations leading to
	faster and expanded flow and increased inter-particle friction coefficient at
	large lubricant contents leading to slower and dense flows. The damping effect
	progressively increases such that at highest concentration studied the flow
	rates reduce than the base case value. It can, thus, be expected that the flow
	will completely cease to stop at very high lubricant concentrations due to very
	high friction coefficient not possible to overcome by induced gravitational
	force for initiating the flow.
	
	We now revert to fig.~\ref{bulkflowsys}b to discuss the lubricant affected flow
	behavior in concentric cylinder system. The improved material flowability, due
	to reduced inter-particle friction, is reflected in terms of reduced torque at
	intermediate lubricant concentrations similar to that observed
	previously~\cite{navaneethan05,wang17} . The opposite behavior is seen for
	increased lubricant concentration. While these observations are qualitatively
	similar to those seen in chute systems (fig.~\ref{bulkflowsys}a) though there
	are certain quantitative differences. In spite of being confined in the shear
	direction, the relative improvement in flowability (reduction in torque) at
	intermediate concentrations is much higher, almost similar to the unconfined
	chute system. Secondly, the decrease of the torque to the minimum (i.e. maximum
	flowability) is quite gradual and occurs over a wider range of lubricant
	concentrations compared to that for chute system which occurs over a much
	smaller range. Further, at higher lubricant concentrations, the flowability
	does not reduce as dramatically as in chute system. In fact, for the entire
	range of lubricant concentration studied, the flowability never worsens than
	the base case. The observed qualitative as well as quantitative differences
	between the lubrication effects seen in the two flow systems can arise due to
	different variables governing the flow occurrence, which we discuss below.
	
	One of the key differences is the substantially increased time duration of
	shear in concentric cylindrical system as against very limited duration in
	chute system. The increased duration of shear, most likely, allows the
	lubricant particles to stretch and smear over a larger surface area of the bulk
	particles, thereby reducing the inter-particle friction, hence improved
	flowability. Indeed, such dependence on time duration has been observed
	previously for very fine pharmaceutical powders~\cite{zhou2011a,wang17} mixed
	with lubricant in a system having similar flow behavior to that in concentric
	cylinder system. The improved flowability was attributed to reduced
	inter-particle cohesion for such fine powders as against the larger particles
	studied over here, wherein cohesion is nearly absent and reduced inter-particle
	friction is the primary driving force behind improved flowability. The results,
	then, also indicate the universality of MgSt particles to modify inter-particle
	interaction, be it cohesion in fine powders or friction in larger particles.
	Further, this effect of time duration gets accentuated due to highly confined
	nature of the flow in concentric cylindrical system. This can result in
	sustained contacts between lubricant and bulk particles as against fleeting
	contacts between agitated and colliding particles in chute system. The
	sustained contacts may aid the lubrication spreading over the bulk particle,
	thereby improving flowability. The relatively quiescent presence of bulk
	particles in concentric cylinder system versus the colliding and agitated
	particles is the result of shear stress in the latter exerting across most of
	the flowing layer as against that exerted over a very small region near the
	central rotating cylinder in the latter case. The quiescent or very slowly
	moving particles may aid in increased spreading of lubricant over the bulk
	particle surface.  The measured velocity and area fraction profiles obtained by
	visualising the free surface of the shearing bed did not reveal anything
	specific to support the above conjecture, hence are not included over here. The
	reason for this, perhaps, is that the flow at the free surface does not
	adequately represent the (confined) flow present at different heights below the
	free surface, while the torque measurement represents the force encountered for
	the entire bed.
	
	\subsection{Lubricant effect during flow of different particle types}
	\label{sec:lubr-effect-part}
	
	Having discussed the shear and confinement effects on the lubricant
	effectiveness to alter the inter-particle friction, we now discuss how the
	presence of lubricant influences the flow of particles of different surface
	morphology in each of the shear geometries. We would like to stress that the
	primary focus here is to understand relative change for each particle with
	respect to the base case, i.e. absence of lubricant.  As mentioned in previous
	section, other physical properties of the materials like density, contact
	mechanics and packing structure may also affect overall flow, the contribution
	of which cannot be ascertained easily. Thus, exact comparison of flow behavior
	across particle types is not possible. Given the opaque and non-reflecting
	nature of particles other than glass beads, particle identification, hence flow
	measurement is quite difficult which gets accentuated with lubricant powder
	addition. Thus, flow profile analysis was not carried out in these cases and
	further discussion in this section is focused only on the bulk flow
	measurements.
	
	\begin{figure} \centering \includegraphics[width=0.8\textwidth]{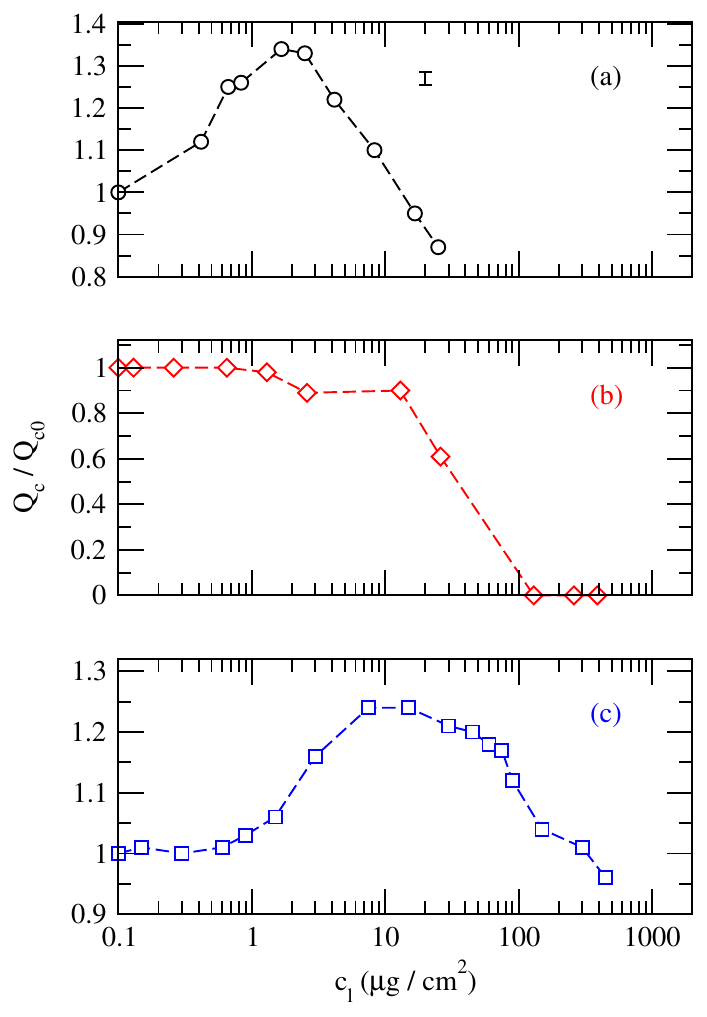}
		\caption{Variation of normalised volumetric flow rate ($Q_c$/$Q_{c0}$) with
			lubricant concentration ($c_l$) in the chute system for (a) Glass beads (b)
			Mustard seeds and (c) Lactose granules. Single error bar shown in (a)
			represents the maximum deviation} \label{chute-particles} \end{figure}
	
	Figure~\ref{chute-particles} shows the variation of normalised volumetric flow
	rate ($Q_c$/$Q_{c0}$) of different bulk powder materials with varying lubricant
	concentrations in the chute system.  Qualitatively, the profiles pretty much
	mimic the variation of effective friction coefficient with lubricant
	concentration shown in fig.~\ref{friction}. This is not surprising given
	similar flow characteristics in chute system and the heap flow (for static
	angle measurements), viz. free surface flow with enough flexibility for flow
	expansion and limited duration of shear. Quite obviously, mustard seeds show no
	improvement in flow rate at all concentrations, while Lactose particles and
	glass beads show considerable improvement in the flow rate at intermediate
	lubricant concentrations, with Lactose exhibiting enhanced flow rate over a
	much wider range of lubricant concentration. In all cases, the flow rate falls
	below the base case for large enough lubricant content. The overall behavior
	owes its origin to the non-monotonic dependence of effective friction on
	lubricant concentration which influences the flow rate variation. The flow
	geometry seems to impose no additional effect on the overall behavior. While
	not measured, the particle level data for Mustard seeds, and Lactose particles
	can be expected to show behavior similar to that obtained for glass beads
	comprising flow expansion and contraction and flow quickening and slowing (see
	fig.~\ref{chuteflowdetails}).
	
	\begin{figure} \centering \includegraphics[width=0.8\textwidth]{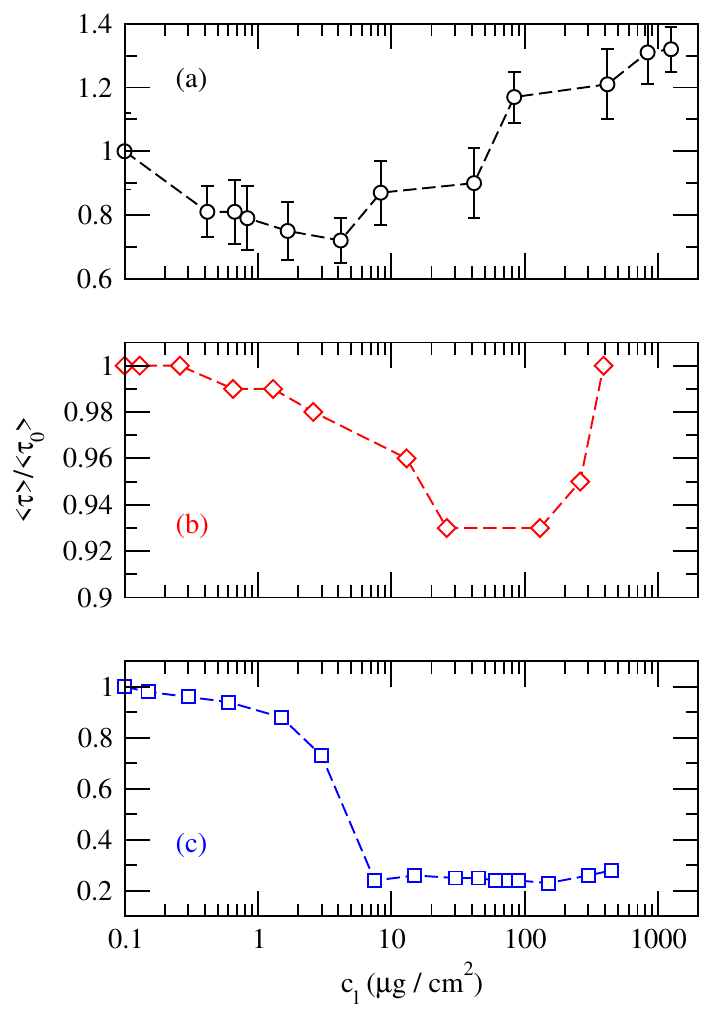}
		\caption{Variation of mean normalised torque ($\langle \tau \rangle/\langle
			\tau_0 \rangle$) with lubricant concentration ($c_l$) in the concentric
			cylinder system for (a) Glass beads (b) Mustard seeds and (c) Lactose granules
			for inner cylinder rotational speed of $5$ rpm.} \label{concyl-particles}
	\end{figure}
	
	We, next, discuss the effect of lubricant on the flow of three particles, but
	now in concentric cylinder system. The variation of average normalised torque
	($\langle \tau \rangle/\langle \tau_0 \rangle$) is shown in
	fig.~\ref{concyl-particles} for the three particle types. The addition of
	lubricant improves flowability for all the three particles, though the maximum
	improvement with respect to base case is seen for Lactose granules while
	minimum improvement is observed for Mustard seeds. This is qualitatively in
	line with friction measurements for each particle described earlier.
	Interestingly, the mustard seeds do show improved flowability in contrast to
	its near inertness with the lubricant content exhibited in terms of friction
	coefficient (fig.~\ref{friction}b and chute flow~\ref{chute-particles}b). This
	may be attributed to the improved level of coating possible due to shearing
	over a much longer duration in concentric cylinder system. The peculiar
	affinity of Lactose is evident in fig.~\ref{concyl-particles}c wherein for the
	range of lubricant concentrations studied, the torque never rises, once it is
	reduced to minimum. The flowability can be expected to worsen in this case only
	at a very large lubricant concentration (not studied over here).
	
	In all the cases, the maximum improvement in the flowability (or minimum torque
	required) is much more than that achieved in inclined chute flow systems. At
	the same time the lubricant content corresponding to this maximum flowability
	is much larger compared to that observed in inclined chute. As discussed in
	previous section, the improved effect of lubricant on the flowability can be
	attributed to increased shearing duration ($\sim 1$ hour), higher confinement,
	sustained contacts, relatively quiescent system lesser agitation present in
	concentric cylinder system as against the lower shear duration (few seconds),
	fleeting contacts, unconfined and collisional flow observed in chute systems.
	The combined effect of these difference is to increase the surface area of bulk
	particles coated by lubricant, thereby reducing inter-particle friction and
	improved flowability.
	
	\section{\label{sum}Summary}
	
	To summarise, we have experimentally studied the effect of tiny powdered
	lubricant on the flow of larger powder particles of different surface
	morphologies in two different flow geometries. The experimental measurements
	comprise those carried out at the macroscopic (or bulk) level like flow rates
	and torque as well as those carried out at the microscopic (or particle) level
	like profiles of velocity, volume (or area) fraction and flow thickness. In
	each experiment, a predetermined amount of Magnesium Stearate (MgSt), as a
	powder lubricant, was mixed with the bulk powder particles and flow of the
	mixture was studied in different shear systems.
	
	The FE-SEM imaging of individual bulk particles, before and after lubricant
	addition, indicates that the inter-particle friction coefficient is primarily
	reduced due to accumulation of lubricant on the particle surfaces. This reduces
	the inter-particle friction at contact for all types of particles. However, for
	particles with rough, uneven texture as for Mustard seeds or Lactose granules,
	there is a possibility for the lubricant to fill in the gaps and reduce the
	unevenness which can enhance reduction of inter-particle friction, though this
	mechanism could not be verified independently. In either scenario, addition of
	lubricant in small quantities reduces the inter-particle friction while
	addition beyond a certain concentration leads to uneven coating of lubricant
	thereby increasing the roughness and the inter-particle friction than that in
	absence of lubricant. The relative reduction in the inter-particle friction of
	bulk powder particles and its dependency on the lubricant concentration
	exhibits a strong dependence on the type of particle surface morphology.
	
	When subjected to different flow conditions, the lubricant particle mixture
	exhibits differences in flow enhancements over the same range of lubricant
	concentration. The effect due to the type of flow, thus, superimposes over the
	altered inter-particle friction due to the added lubricant. The non-monotonic
	dependence on lubricant concentration, i.e., improved flowability at
	intermediate concentrations and worsened flowability at very large
	concentrations, is, however, preserved in most cases. The increase in the flow
	rate in the chute system is a result of faster particle motion due to reduced
	inter-particle friction and flow expansion, resulting in lower particle
	fraction and increased flow thickness. The exact opposite behavior, viz.,
	compressed flow with higher particle fraction and slower particle motion, is
	observed for higher lubricant concentration. However, the relative change in
	inter-particle friction in the concentric cylinder geometry is quite
	significant. The reason for this difference is the extensive time duration over
	which the lubricant particle mixture is sheared in the concentric cylinder
	system compared to the very short duration in the chute system. In the former,
	it seems that extensive shearing allows the lubricant powder to coat uniformly
	on the surface of bulk particles, thereby extending the effect of the
	lubricant, which seems difficult to achieve over a short duration of shearing
	in the inclined chute system. Even the particles with the most surface
	roughness seem to be lubricated quite well in the concentric cylinder system,
	which is, otherwise, not possible in the chute flow system.
	
	The extensive study in this work builds significantly upon the outcome from our
	previous work~\cite{ghodake2022flow}. While the previous study provided early
	results of the lubricant effect for one type of particles and shearing system,
	the current work expands and validates the applicability of lubrication across
	particles of different surface texture and different shearing systems. It is to
	be noted that the predominant qualitative dependence of flow on lubricant
	concentration is preserved for different cases, suggestive of near universality
	of the observations. However, certain nuances are still observed. While, the
	duration of shearing shows substantial influence on the lubrication effect,
	presumably due to ability of lubricant to spread effectively over particle
	surface, the effect of confinement remains muted. Perhaps, experiments need to
	be carried out in a confined system with short flow duration, as in a silo
	system to better understand the effect. The nature and texture of the particle
	surface does have a clear influence on the lubrication effect suggesting of a
	peculiar affinity with lubricant particle type. However, we are unable to show
	this effect through additional quantitative estimates. Clearly, the combined
	effect of shear type and particle surface produces different lubrication
	effects for the same particle type and same lubricant content per unit area.
	
	Overall, the effectiveness of the lubricant in enhancing the flowability is
	quite sensitive to the type of flow application wherein the lubricant coated
	material is being processed and the system handling such mixtures needs to be
	designed accordingly. Given that the two systems considered over here will
	comprise most shear conditions encountered in most applications, these
	observations can be considered as a basis for designing applications handling
	lubricant-particle mixtures. Primarily, the lubricant concentration needs to be
	carefully adjusted with respect to the type of shearing geometry employed.
	Having said that, it would be interesting to obtain a detailed investigation of
	the lubricant coated particle surface, perhaps using an innovative surface
	profiler which is deferred to a later study.
	
	\section*{Acknowledgements} The authors are grateful to Dr. Pankaj Doshi from
	Pfizer Products India Private Limited for several fruitful discussions and his
	insightful suggestions. AVO gratefully acknowledges the financial support from
	Science \& Engineering Research Board, India (Grant No. CRG/2019/000423).
	
	\bibliography{References}
	
\end{document}